# The Transformative Impact of AI and Deep Learning in Business: A Literature Review

Dias, Fabio S.  Lauretta, Grace A.

*Abstract* — **This paper aims to review the radical role of AI and deep learning in various functional areas of the business, such as marketing, finance, operations, human resources and customer service. Thus, based on the overview of the latest research and practices focusing on AI technologies in different industries, the possibilities of improving organizational efficiency by personalized AI for making decisions based on big data and personalizing clients' interactions with organizations are presented and discussed. Several operational issues, ethical concerns, and regulatory concerns have also been discussed in the review of the literature. Moreover, it covers material applications in the healthcare sector, the retail and manufacturing industry, agriculture and farming, and finance before considering possible future developments and themes for further investigation. Drawing from this revolutionary ethnographic review, organizations aiming to implement strategic and responsible optimization benefit from detailed guides.**

## I. INTRODUCTION

Artificial Intelligence (AI) and deep learning are beginning to revolutionize many business processes across the fields of decision-making, business process optimization, and customer relations. This literature review seeks to highlight the effects of AI on business, with specific emphasis on the various sectors in which it is applied and the advantages and existence of the technology. This research paper aims to provide a fair outlook on the use of AI in business through a review of the current research and prevailing practices.

*A. Purpose of the Article:*

The purpose of this literature review is to synthesize the literature in order to provide an understanding of how AI and deep learning can support business processes. This review conceptualizes the application of AI within the context of effectiveness and explores the innovation and competencies available within different domains. The review will consider the benefits of AI implementation, such as automation, predictive analysis, and better customer understanding. There are threats like data quality, ethical issues, and regulations. Awareness of AI's opportunities and risks enables business managers to decide on implementing these tools and consider the advantages derived from it against its installations and implementations [1].

*B. Relevance of AI in Business:*

AI's role is significant across any business field, which includes marketing, finance, logistics, and involving customer relations. With an increasing volume of data accumulating in organizations, business AI applications allow for better recognition and decision-making [2] [8]. Some examples of marketing and customer analytics, scam prevention and detection in the finance sector and predicting equipment failures in manufacturing industries show how flexible and productive AI proved to be. With mounting investment in developing AI structures, firms worldwide have been perceiving the value of incorporating AI to sustain competitive advantage.

*C. Key Concepts:*

In order to understand AI's specializedness, let's distinguish between AI/ML/DL. AI analyzes use technologies that permit human-like conception or decision-making, comprehension, and learning. Artificial intelligence, or AI, is a larger category but one specific discipline that specifically deals with systems that learn over some time and get better at identifying data patterns. Computational learning methods have come into play to improve machine learning with tools known as Deep Learning Neural Networks that analyze various structures of data. These differences explain the nature of additional business applications of AI on the basis of which ML and DL act as tools of AI [3] [4].

*D. Scope and Structure:*

This paper discusses the basics and recent developments in AI and DL, the current uses of AI across business functions and prospects, current AI issues and possibilities and other ethical concerns. This is discussed under the subtopics of the theoretical framework and literature review involving a detailed look at how AI supports marketing, financial, Human Resources, supply chain and customer service departments. The review also ends with a focus on different future AI trends and research directions. Through this structure, readers are given sufficient knowledge on how AI can revolutionize business operations and helps readers prepare for the keynotes important to AI strategy and its implementation [5].

## II. BACKGROUND AND THEORETICAL FOUNDATIONS OF AI AND DEEP LEARNING

The incorporation of AI in businesses has gone through different phases with some underlying reasons for its growth. In this section, the authors present the development of AI concepts, technology fundamentals for deep learning, and the importance of big data and computational capabilities in today's AI applications.

### A. History and Evolution of AI in Business:

The deployment of AI into business occurred in the last two decades of the twentieth century, while the basic work that prepared the ground for what later became the AI industry was initiated at the beginning of the decade of 1970 with expert systems and decision support systems. Among the first domains of its application were credit assessment and the use of measures to combat fraud, and both took place in the eighties and nineties. It was the Internet and the stoke of data, which became possible to millions through the nineties, which really pushed the advance of AI and created an environment for business holding [6]. Some of the major events of change involved included, for example, the improvement of Google Search algorithms on Search Engine Optimization scavenging on principles of artificial intelligence as well as the improvement of Machine Learning Techniques or methods for carrying out Predictive Analysis. Artificial neural networks became a groundbreaking technology around 2010 to assist companies with employing neural networks for the sake of an image recognizer, a language processor, or a data predictor.

Today, AI applications span various business domains. In marketing, AI adds contextual and behavioral aspects to recommendation engines and real-time ad inventory bidding, and in finance, algorithmic trading and credit risk assessment. The present wave of employing AI is a result of recent trends in using data analytics in business settings to enhance organizational performance, revenue, and customer engagement.

### B. Technological Foundations:

AI's essence is to include methods that imitate specific human processes in the mind to correspond to patterns and make decisions. Deep learning, being a subset of the machine learning concept, encompasses the artificial neural networks imitating the structures and workings of the human brain. A neural network, as the name suggests, is constituted by nodes or neurons that are connected in the form of layers, and the data is processed time and again by the layers to predict with high probability [7] [9].

Deep learning techniques are typically categorized by their learning methods: classification learning and regression learning, along with supervised, unsupervised, and reinforcement learning. They supervised learning trains on labeled data where the algorithms learn to predict or categorize into known classes. In contrast, unsupervised learning comprises a pattern analysis of data without their labels; it can, therefore, be applied to tasks such as clustering customers and identifying outliers cases [10].

### C. Importance of Big Data and Computational Power

The coming of big data and better computer hardware has played a key role in the growth of AI and deep learning applications. Big data is all the structured and unstructured information emanating from sources such as social media, transactions, and sensors. The availability of large datasets allows AI models to identify patterns and correlations that would be challenging for traditional data analysis techniques [8] [11]. Businesses rely on big data to fuel AI systems that generate insights from customer behavior, supply chain performance, and financial markets.

Deep learning has also been influenced in large measure by the co-occident development of computing power, particularly graphical processing units (GPUs). GPUs are always employed for data flows and training multi-layered architectures, and they contribute to reducing the time for AI training. As of today, there are hundreds of platforms offered by various companies, such as Amazon Web Services or Google Cloud, that ensure AI development and offer businesses various resources on various levels. They facilitate the collection and processing of large amounts in business and give more people access to AI applications .

This development has, therefore, brought about challenges regarding what big data and other computational accomplishments have done in expanding the scope of AI. They have pushed it to solve many issues facing businesses. They also free organizations from the constraints of traditional business analysis to work and learn in real-time, creating new horizons of predictive and prescriptive possibilities for organizations.

## III. LITERATURE REVIEW: CURRENT APPLICATIONS OF AI AND DEEP LEARNING IN BUSINESS

The application of AI and deep learning across various industries has led to innovative changes from countries to corporate organizations and individual companies, translating the topic into numerous documents. This section also presents selected papers that examine how AI is being applied in particular business processes, explaining specific industries where AI has made dramatic changes. These applications range from marketing functions, finance functions, supply chain management, human resources, customer service, product development, and sector-specific applications like health care, retailing, manufacturing, agriculture, and finance industries.

A. *Introduction to the Literature Review:*

This study aimed to review the existing literature systematically in order to assess AI/Deep Learning in business organizations and its prospects, carefully choosing trustworthy academic databases, such as JSTOR, IEEE Xplore, Google Scholar, etc. Some of the terms used in the search were AI and Business, deep learning use cases, sectorial AI, and corporate use of Machine learning. These keywords were useful in a specific search that made sure that the selected papers covered the controversial change of AI in different businesses and industries. The screening criteria for the articles focused on AI technologies in the last decade, particularly starting from the year 2015 and above; works from industry reports, international, peer-reviewed journals, and leading conferences with papers on AI.

The works surveyed are to give an account of a diverse range of fields such as finance, healthcare, retail, manufacturing, agriculture, logistics, and other industries because AI and Deep Learning are used as universal tools in a data-driven world. By location, emphasis was placed on investigations from North America, Europe, and Asia since these areas are most active in AI investment and infrastructure. At the same time, North America led in regulations, Europe followed, and Asia was quickly improving in artificial intelligence applications, especially in customer relations and production.

The review, analyzing a variety of industries and geographical areas, provides an overall understanding of how AI deals with the specific issues of enterprises from different fields and geographical locations, from customer orientation in retail to using AI for predicting failures in producing industries. A vast amount of the literature also provides similar findings [15].

B. *AI in Different Business Functions:*

The use of artificial intelligence (AI) across business functions has greatly interrupted real working processes by increasing effectiveness and precision and providing insightful approaches. The present section discusses how AI is being implemented within major organizational fields based on research, surveys, and relevant works as well as case studies to highlight how AI is affecting these key organizational functions.

- Marketing and Sales:

Self-service tools through artificial intelligence have now become strategic and even revolutionary in marketing and sales since data is big and complicated. Customer profiling, recommendation engines, target marketing, and product recommendations based on a detailed understanding of the customer are possible by using AI. Recommendation systems are popular with most contemporary firms, such as Amazon and Netflix, in which recommendations are generated from user preferences and behaviors, improve customer experience, and increase loyalty [4] [14] [18].

Research shows that branded messaging endorsed through artificial intelligence is considerably more efficacious than eclectic mass marketing. Probability algorithms such as those based on AI, for instance, can help determine the future conduct of the client depending on the previous exchanges to enable marketers to customize the content and time of delivery besides platforms. According to studies, businesses that use AI in the promotion of their products or services record a better ROI due to improved CTR and customer loyalty. Also, the case studies show that Amazon's adaptation of the recommendation system contributes to about half its revenues by using artificial intelligence to recommend products similar to those the customer has been in the habit of buying [17].

- Finance and Accounting:

The finance sector has had a good uptake of the technology to look for fraud, credit worthiness, trading, and in looking at the future. The specific AI fraud detection systems analyze transaction-related data and check for patterns that may be part of fraud prior to its escalation [6] [7]. Manufacturing needs safer and more secure supply chains by using machine learning algorithm-based solutions like the services offered by JP Morgan Chase & Co for identifying frauds and Wells Fargo & Co for real-time, AI model-based monitoring of behavior to quickly and efficiently lower response duration and loss.

AI has introduced some social transformations in credit scoring, such as where lenders assess loan applicants and how they assess them. The conventional models meet this requirement by including only a limited number of parameters. In contrast, many AI credit models, including the one adopted by ZestFinance, have included many parameters, such as transaction histories and social and spending patterns [11] [19].

AI has taken automated trading to another level because it means a machine can make trading decisions involving such large segments of data in a few milliseconds at most. In addition, in the fields of financial auditing and compliance with essential significance in business, AI tools like PwC's GL.ai are applied to find out the specifics in a large amount of analyzed data and decrease the time and the number of audits.

- Supply Chain and Operations:

AI applications in supply chain management include Predictive maintenance, Logistics optimization, and Demand forecasting. It is especially important in sectors where equipment plays a vital role, including production as well as aerospace. From the data of the sensors, AI algorithms can determine when the equipment would likely fail. Thus, it is repaired ahead of time, cutting on working time loss. General Electric (GE) is an example of a firm that employs AI in predictive maintenance, cutting costs because the firm can anticipate equipment failures before they cause downtime [9].

Another area of significant advancement is logistics optimization. AI models use characteristics such as traffic, weather, and demand to make relevant estimations on the best delivery routes that will minimize fuel usage while enhancing delivery times. Some examples include DHL and FedEx, where the companies rely on AI logistics that display the best routes and arrive closer to time, as one can imagine, resulting in better customer satisfaction [16] [21].

Ultimately, businesses can manage their inventories well due to this capability, which prevents instances of overstocking or stockouts. By synchronizing supply with demand expectations, the value chain is optimized, costs for products and services are decreased, and thus, customer satisfaction is enhanced.

- Human Resources and Recruitment:

AI is used in human resources (HR) for talent acquisition and management, retention, and performance appraisal [12]. Online today, the Internet is used to source for employees, and this involves the use of AI systems that assist in filtering resumes so that they match the job descriptions. LinkedIn, for example, uses artificial intelligence to match suitable candidates with job profiles based on their qualifications, skills, and experience, as well as their likelihood of fitting into an organization's culture, thus improving recruitment and reducing the time to hire [5].

Implementations of artificial intelligence in the management of employees for purposes of attracting, keeping, and later on satisfying the employees also receive attention. Presently, organizations can apply predictive analytics within the Human Resources department to identify staff considered high-risk because they have diminishing levels of performance, satisfaction, or frequent voluntary turnover. Evidence from the information is that an HR department may be useful in the prevention of issues that cause a loss in morale and turnover. However, there is one social concern when it comes to the integration of AI in the HR field: bias [17]. Another study goes further to unveil that if it is properly utilized, the historical data will have an impact on hiring bias representation, and it also shows prejudice in the hiring model.

- Customer Service and Support:

AI has given so many advantages to customer support services such as chatbots, virtual assistants, and sentiment analysis. Customer support personnel who integrate NLP-based chatbots can, in real time, answer a number of questions that customers may pose [10]. Customers of companies like H&M and Sephora, for instance, also use chatbots to assist them with shopping and orders. These avatars must be capable of responding to other basic queries, thereby relieving human beings who always address the call center of such tasks. There are always improved response times when using Artificial intelligence in customer servicing because it offers fast and efficient problem-solving [14].

Most big and small businesses use sentiment analysis programs to monitor their ratings, positive and negative comments, and feedback received from the public through social media and other platforms. Through AI models, customers' attitudes are recognized, which in turn enables firms to solve certain problems before they arise and adjust their business approaches correspondingly [15].

- Product development and Innovation:

AI is also instrumental in product development and as an innovation enabler in industries where value-chain research and development is long, including healthcare and tech. For instance, AI apps in drug discovery reduce the time spent on research by forecasting intermolecular relations and results. Business organizations, including Pfizer and Novartis, apply artificial intelligence to model and anticipate responses to a selection of different drug combinations before they are actually tried out on patients [19].

In the tech industry, design and prototyping are accelerated by artificial intelligence. True engineers and designers involved in the design of products perfect by trial and error make use of software that gives quick solutions by producing a number of versions that are tested several times as compared to the traditional techniques and tools. There are many applications of this capability, particularly in the context of product design and testing where time is of the essence [22] [23].

Besides, applications of AI are present in market research and customer customer feedback analysis that help companies identify their unique selling points. For instance, machine learning recognizes customer opinions that are similar to one another so that companies can improve their products; similarly, in the automotive industry, product design applications that use AI help in designing prototypes of vehicles, thereby making the designing process fast and offering better products .

*C. Industry-Specific AI Applications:*

- Healthcare:

In terms of AI's influence on the healthcare system, the three areas are diagnostics, individualized treatment, and administrative support. Hence, diagnostic tools supported by AI help diagnose ailments through medical imaging early enough, such as cancer. Clinical decision support in omics greatly enhances patient outcomes because treatments are based on personalized patient information. Also, AI simplifies processes involved in record keeping and billing, increasing productivity and decreasing healthcare workers' workload [19].

- Retail:

The application of AI in retail involves the list of areas specific to retail business, such as inventory control, demand forecasting, and customization of clients. In managing inventories, AI makes predictions on the needed stock to order in order to avoid overstocking as well as stockout. Many large companies in the e-commerce industry apply AI for the recommendation of products, which ultimately increases their sales and the satisfaction of their clients [10].

- Manufacturing:

AI facilitates quality control, robotics, and predictive maintenance in manufacturing. AI-driven quality assurance techniques identify potential imperfections in a product at the same time, thus increasing product quality. In robotics, artificial intelligence helps in decision-making, which gives little or no room for human involvement in a number of activities. Predictive maintenance also enriches productivity by reducing the number of surprises associated with unfavorable failures of equipment and lengthening the lifetime of the machinery [9].

- Agriculture:

AI has become an important tool in precision farming, enabling the observation and management of crops, soils, and other resources. Artificial intelligence drones provide crop conditions, and machine learning provides better water and fertilizer use to increase yield and resource efficiency. The use of AI in agriculture enables farmers to optimize their operations, cut down on expenses, and improve decision-making.

- Finance and Banking:

AI applications in banking encompass customer analytics, compliance, and risk management. For example, AI models analyze customers' data for fraudulent activities so that banks adhere to regulatory requirements and manage financial risks. These apps make banking safer and more effective and improve customer experience through customization services [19].

IV. CHALLENGES AND ETHICAL CONSIDERATIONS IN AI ADOPTION FOR BUSINESS

The current literature about AI and deep learning applications in business highlights the wide range of roles of these technologies. For each function and industry, AI has unique benefits that enable businesses to innovate, streamline operations, and make better decisions. Null In this part of the discussion, we examine the concerns, including these topics, constraints, and legal issues that come with using AI in business environments in more detail.

*A. Technical and Operational Challenges:*

One of the issues in the deployment of AI technology is quality maintenance and data fusion since the tools of AI are tightly coupled with the requirement of vast amounts of quality data to produce reliable information. Data available to businesses may need to be more cohesive and consistent or contain errors, which can hinder the accuracy of AI models. Scalability is another major concern, wherein AI models that perform satisfactorily on small datasets are likely to perform poorly as the increased data scale, requiring more powerful hardware and more storage, may be quite challenging [6] [8].

Further issues are maintenance and updating AI models. AI technologies, particularly deep learning models, have to be updated regularly so that they remain competitive and up to date. This process can be costly and time-consuming, as it often involves retraining models and incorporating new data.

*B. Ethical and Social Concerns:*

There are many ethical implications when it comes to AI concerns, including possible bias within AI algorithms. The reason may either be the training data employed or the model itself has design defects. One of the issues of concern is the fact that algorithms are biased, which results in the discrimination of some groups of people in sensitive issues such as recruitment, credit scoring, or even law enforcement [8].

In modern society, AI also poses a threat to people by unbalancing the overall activity of the workforce. Automation through AI can lead to job displacement, especially for roles that involve repetitive or predictable tasks. While some argue that AI will create new opportunities, studies show that automation can lead to job insecurity and reduced job availability in certain industries. CEOs are thus under pressure to pursue digital progress in a socially responsible manner and offer solutions for the transformation of the workforce that very much involves retraining and upskilling workers.

Privacy is another concern that grows more apparent as more and more applications make use of user data to deliver tailored services. Consumers sometimes require being informed on the way their data is collected and utilized, an element that brings issues concerning misuse of data. Current rules, including GDPR, EU, and CCPA, entail strict policies on how data may be used in AI with firm guidelines on how businesses should deploy the technology [13] [21].

### C. Regulatory and Legal Aspects:

Legal regulation of AI is in progress as government and international organizations aim to close the gap between trends and laws. Proposed rules such as GDPR and CCPA are meant to safeguard consumers because they put preeminent restrictions on data acquisition and application. Companies have a responsibility to make certain that their AI systems are law-abiding, or else they will suffer the consequences of the law. Moreover, there is an active debate about the responsibility of AI systems, particularly when they produce some negative results. [12]. The necessity of ethical principles for AI is more appreciated, and guidelines, including the rules by the European Commission and IEEE, put present aspects of impartiality, openness, and responsibility at the forefront. It has been found that more and more companies have started applying AI systems in business, and there is a need to consider these ethical frameworks and the governance structure that will meet the requisite standards of the law [9].

## V. FUTURE TRENDS AND EMERGING RESEARCH IN AI FOR BUSINESS

### A. Advances in Deep Learning Techniques:

The advanced growth of artificial intelligence (AI) and deep learning causes them to give fresh tendencies and fields in the field of AI utilization in business. These are some of the topics discussed in this section: Deep learning techniques, AI convergence, and its ability to support new business setups and facets of research.

New deep learning techniques, such as transfer learning, explainable AI (XAI), and reinforcement learning, have broadened AI's scope in business. Transfer learning enables AI models to be adapted for new tasks with minimal additional training. This proves especially useful in niche applications where data is limited, such as specialty healthcare diagnostics. Explainable AI addresses the "black box" nature of deep learning by making model decisions more interpretable, essential for sectors like finance and healthcare, where transparency and trust are paramount [7].

Reinforcement learning, a technique in which AI models learn through rewards and penalties, is increasingly applied in dynamic environments. For instance, in supply chain management, reinforcement learning enables a business to manage its supply chain inventory and logistics more responsively in the modern climate and conditions for buying and selling [22].

### B. Integration of AI with Other Technologies:

AI's connection with IoT, blockchain, and edge computing extends AI's business uses even more. IoT involves collecting data from connected devices and channeling the same information to AI in an effort to monitor assets such as factory machines or logistics vehicles. For instance, IoT and AI may be applied in manufacturing to forecast equipment maintenance requirements and reduce overall time for repairs and expenses [10].

Blockchain and AI have traditionally been functionalities that work well together. Blockchain ensures security and transparency in AI's data-processing roles, especially in the financial and supply chain industries. Blockchain offers an added layer of security by recording data used in AI training, reducing data manipulation risks, and ensuring traceable, reliable insights.

Edge computing, which processes data closer to its source rather than through centralized cloud systems, benefits industries requiring real-time responses, such as autonomous driving and industrial automation. By enabling local data processing, edge computing reduces latency, making AI-driven decisions faster and more efficient [23].

### C. Potential of AI in New Business Models:

AI is also driving new business models, especially within the gig and shared economy. In the gig economy, AI optimizes the matching of freelancers with clients by analyzing project needs and skills. It should be noted that AI has an essential role in platform-based organizations such as Uber or Airbnb, thereby helping to determine variable prices, offer the most relevant suggestions, and allocate resources according to need, all of which contribute to improving customer experiences and improving operational experiences [14].

It established that a shared economy platform uses AI to analyze usage patterns of the resources in a platform and hence enhance resource utilization together with customer satisfaction. These applications demonstrate how AI can be used to enhance service delivery and provide individually tailored first contact for users, which would feed into the value factor for competitive industries [15].

*D. Ongoing Research Directions:*

Even in the current era, there remain places that need more improvement in terms of research, the biggest of them being ethical AI. Another important area of research is the so-called 'regulatory compliance' that navigates to decisions in fair and legal manners. Other interesting and developing AI architectures are the so-called hybrid AI systems that include the use of machine learning associated with rules. Such a system enables business organizations to align AI solutions with their contexts more flexibly and responsively [18] [20].

Other emerging research topics include the development of AI for industries because each has distinctive problems, such as those in agriculture, forestry, or environmental science. Industry-specific solutions are often more profound and more accurate than general solutions since they are based on specific information.

## VI. CONCLUSION AND IMPLICATIONS FOR BUSINESS LEADERS

Knowing this, this review has highlighted how AI and deep learning are transforming business, with a focus on numerous uses across marketing, finance, operation, and customer service. In predictive analytics, through customer interaction decisions, AI has impacted business tremendously, its decisions, and customers.

*A. Summary of Key Findings:*

The implementation of artificial intelligence in business highly influences its work since it executes tasks that could hardly be completed or were impossible to complete when using simple tools. The first use of AI in marketing is quite evident from the recommendation systems used to drive up conversion and dynamic pricing. They apply AI for identity checks and credit checks, which is basically an estimate of the candidate's creditworthiness, in bot trading that brings in security while at the same time delivering optimal results. MCP managers from the supply chain and operations apply it to areas such as predictive maintenance and logistics, whereas HR MCP managers use the tool in recruitment and engagement programs [15].

However, the challenges of deploying artificial intelligence are as follows: Issues like data quality, data integration, issues of scale, and finally, issues inherent to the maintenance of a model can sap the motivation of practitioners. However, there are a few other ethical concerns, such as machine learning algorithms, privacy invasion, and more.

*B. Practical Recommendations for Businesses:*

A structured approach is essential for businesses considering AI adoption. One major factor managers should look at is the use of the pilot of the AI for a particular function in order to assess the idea before replicating it across the firm. The major here with regards to AI is to set direction on the data since quality and availability are among the top influential features of a model. To make the models work effectively, organizations have to invest in cleaning, blending, and warehousing the data to feed the artificial models well [19]. Even more, the internal capacity to develop or buy can have a positive impact on AI deployment since such extensions may demand technical abilities for model changes. The following recommendations are also necessary: Ethical considerations into AI strategies, guidelines for monitoring and mitigating bias in machine learning and AI roadmaps, safeguarding of customers' data information, and clarity concerning AI-related decision-making processes.

*C. Final Thoughts on the Future of AI in Business:*

Far from being limited to trends at the moment, it is safe to say that in the future, AI will take up an even larger proportion of business processes. However, it is the safe and ethical type of adoption that should be promoted. They should ensure that they are aware of current technologies and changes in the law so as to educate their organizations on the best trends. Taking this into account, more companies consider AI as a beneficial weapon for development that can solve emergent needs in the marketplace and enhance the delivery of business value. Thus, even today, AI has produced considerable opportunities for the scale-up and advancement of business activities.